# Practical Aspects of LTE Network Design and Deployment

Ayman Elnashar

## Abstract

*This paper covers the practical aspects of commercial long term evolution (LTE) network design and deployment. The end-to-end architecture of the LTE network and different deployment scenarios are presented. Moreover, the LTE coverage and link budget aspects are discussed in details. Theoretical and practical throughputs of LTE system are analyzed. In addition, capacity dimensioning of LTE system is explained in details and compared with the evolved high-speed packet access (HSPA+) system. Additionally, the quality of service (QoS) of the LTE system and the end-to-end implementation scenarios along with testing results are presented. Finally, the latency of the LTE system is analyzed and compared with the HSPA+ system. This paper can be used as a reference for best practices in LTE network design and deployment.*
**Keywords:** LTE, OFDM, OFDMA, SC-FDMA, HSPA+, WCDMA, latency, QoS.

## I. Introduction

The LTE is the next major evolution step in mobile radio communication, introduced by 3GPP in Release 8 and provides initially downlink (DL) peak data rates of 100 Mbps, an uplink (UL) data rate of 50 Mbps compared to 42Mbps/11Mbps (DL/UL) peak throughputs of HSPA+ R8 3GPP standard [1]-[9]. The LTE system brings flat network architecture with improved data rates and less latency that provide best mobile data browsing experience [2]-[6]. The LTE system brings two new multiple access techniques: orthogonal frequency division multiple access (OFDMA) on the DL and the single carrier frequency division multiple access (SC-FDMA) on the UL [2]-[6], [10]. Both of these access schemes are based on orthogonal frequency division multiplexing (OFDM) transmission technique. The main advantage of OFDM, as is for SC-FDMA, is its robustness against multipath signal propagation, which makes it suitable for broadband systems, compared to wideband code division multiple access (WCDMA) technique [5], [6]. The SC-FDMA brings additional benefit of low peak-to-average power ratio (PAPR) compared to the OFDMA technique making it suitable for uplink transmission of user equipment (UE) to extend the battery backup time [7].

Even though, the WCDMA system can be extended to a broadband system, its complexity increases where it requires more number of rake receiver fingers since its channel is a frequency selective fading channel [8], [9]. Therefore, extension of WCDMA system with high speed packet access (HSPA) evolution to a 20 MHz broadband system requires extension of similar factor on the number of fingers in rake receiver, and thus its complexity [8], [9]. Adding multiple-input multiple-output (MIMO) to the HSPA system on top of the above complexity in receiver design limits the anticipated gain from it [8], [9]. 3GPP adopted other ways of extending HSPA system to broadband systems, based on multi-carrier HSPA with added complexity as well in terms of power amplifier (PA) design, network cost, and network optimization [8], [9]. OFDM can also be viewed as a multi-carrier system but each subcarrier is usually narrow enough that multipath channel response is flat over the individual subcarrier frequency range, i.e. frequency non-selective channel (flat fading) and hence receiver design is very simple. More specifically, OFDM symbol time is much larger than the typical channel dispersion [5]. Hence, OFDM is inherently susceptible to channel dispersion due to multipath propagation. OFDM symbol detection requires that the entire symbol duration be free of interference from its previous symbols i.e., Inter-Symbol-Interference (ISI) spill-over. However, ISI spill-over at the beginning of each symbol can be tackled by adding a cyclic prefix (CP) to each transmit symbol [5].

The aim of this paper is to provide the key practical aspects of design and deployment of a commercial LTE network. The analysis in this paper is presented in a comparative manner with reference to the HSPA+ network to benchmark and evaluate the LTE network performance. The remaining of the paper is organized as follows: Section II provides LTE network architecture and typical implementation scenarios. LTE coverage and link budget (LB) analysis are presented in Section III. Network dimensioning and design exercise are introduced in Section IV. LTE QoS and practical implementation exercises are introduced in Section V. The LTE network latency and a comparison with the HSPA+ network are presented in section VI. Finally, conclusions and key findings are summarized in section VII.

## II. LTE Network Architecture

LTE system brings flat all IP architecture [2], [3], [11], [12]. This flat architecture offers saving in CAPEX and OPEX thanks to eliminating the radio network controller (RNC) and the circuit switch (CS) core while introducing IP multimedia subsystem (IMS) [11], [12]. Moreover, the LTE system offers higher network performance and increased efficiency. This is achieved by reducing the latency since the eNB is directly connected via S1 interface to the EPC and also faster handover thanks to direct connectivity between eNBs via X2 interface. The EPC consists of the serving gateway (S-GW, SGW, or SAE GW), the mobility management entity (MME), and the packed data network (PDN) gateway (PGW) [11], [12]. The SGW is responsible for handovers with neighboring eNB, data transfer in terms of all packets across user plane, mobility anchor to other 3GPP systems (i.e., 2G and 3G). The MME is the centralized control unit for key operations on access network and core network. The PGW is responsible to act as an anchor of mobility between 3GPP and non-3GPP technologies [11], [12]. The PGW provides connectivity from the UE to external packet data networks by being the point of entry/exit of traffic for the UE. In addition, LTE brings the always-on connected concept with less than 100ms transition from idle state to connected state. This concept tackles the impact of the signaling storm generated from smartphones due to the transition from idle state to connected state. Finally, LTE system offers increased throughput with 100Mbps/50Mbps DL/UL peak throughputs with category 3 (i.e., CAT3) modem and with seamless evolution to 150Mbps/75Mbps peak DL/UL throughputs with the introduction of category 4 (CAT 4) modem [2], [7].

A typical commercial LTE network high-level topology is depicted in Figure 2. The network consists of four domains as follows:

1- ***Access Network Domain***: consists of eNBs that provides the evolved universal terrestrial radio access (E-UTRA) user plane and control plane protocol terminations towards the UE and IP access layer that carry the traffic of eNB towards the core network domain. Each eNB may have one S1 connectivity with one EPC node and it can have two S1 connectivities with two EPC nodes if geo-redundancy is adopted similar to the scenario in Figure 2. The S1 consists of user plane and control plane. The S1 user plane (S1-U) is routed to the S-GW and the control plane (S1-MME) is routed to the MME. The control plane capacity is ~ 1%-3% of the user plane capacity. Moreover, X2 connects neighbor eNBs to support UE handover. The X2 capacity is estimated to be around 3-5% of S1 capacity. As per [3], 10ms delay is recommended for S1 and 20ms for X2. Therefore, if the latency requirement is met for S1 then, X2 shall be met accordingly. It is important to mention that, X2 connectivity can be made in the IP security gateway (IPSec GW), which is part of the security domain, or in the IP access routers or even in the IP core cloud. The selection of X2 cross connect node depends on the latency requirement, network topology, and security requirement. The recommended node to terminate the X2 is the IPSec GW to meet security requirement, reduce the latency, and maintain consistent network topology. The S1 latency is divided into two parts; access network delay and transport network delay where the 10ms can be divided 5ms for each network. The maximum latency of the transport network limits the number of cascaded transmission links between the eNB and EPC.
2- ***Core network domain***: This is the main domain that includes IP core network cloud that carries the access network traffic to core network elements. It is to be noted that, the EPC nodes can have local redundancy and/or geo-redundancy for high reliability and high availability. Also, the home subscriber server (HSS) is a new functionality that can be added as a separate entity or via upgrade to the existing home location register (HLR). The service profile of each LTE subscriber is defined in the HSS [11], [12].
3- ***Security domain***: in addition to the well known firewall nodes, two new network elements are introduced as part of the LTE system:
   i. ***IPSec Gateway (IPSec GW)***: it is responsible for encryption and for terminating the IPSec tunnels with eNBs. All traffic and signals are encapsulated inside IPSec tunnel for security purpose and to prevent any attack to the EPC nodes. The deployment of IPSec GW can be in cluster fashion for reliability and simplicity. Therefore, the network can be divided into several clusters (i.e., regions) and eNBs traffic within certain cluster is encapsulated towards an IPSec GW pair with 1+1 hot standby. Furthermore, a feature such as X1-flex can allow the eNB to create two IPSec tunnels with two different IPSec GW nodes in two clusters (i.e., two different regions or locations) for reliability and in this case the traffic can be distributed based on load sharing or hot standby fashions.
   ii. ***Certificate Authority (CA)***: responsible for allowing eNB to access the network by issuing, managing, and validating the digital authentication certificates.
4- ***Operation and management (O&M) domain***: this domain includes O&M functionalities and NMS for all network elements.

### III. LTE Link Budget and Coverage Analysis

The aim of the link budget (LB) is to identify the maximum allowable path loss (MAPL) between the transmitter and receiver for UL and DL and therefore the cell radius can be calculated for different terrain morphologies (i.e., dense urban, urban, suburban, and rural) based on the propagation model. A tuned version of COST231-Hata model is used to estimate the pathloss. LTE Release 8 is a data centric technology; therefore the critical coverage constraint when designing an LTE network would be the expected data rate at cell edge rather than the received signal level. The outcome of the LB calculations enables the network designer to determine the expected coverage calculated in theory and compare it with the measured values in the field. Table 1 provides the theoretical link budgets for typical LTE system at 1800MHz band (i.e., 3GPP band 3) with 20MHz channel bandwidth. The link budgets are calculated at different morphologies and for UL and DL. The cell radius of each terrain is determined based on the smallest cell radius from UL and DL. The theoretical link budgets for all morphologies demonstrate that the LTE system is uplink limited (i.e., MAPL of UL < MAPL of DL) and there is ~3-4 dB between the MAPLs for UL and DL at typical cell edge throughputs (i.e., DL at 512Kbps and UL at 128Kbps).

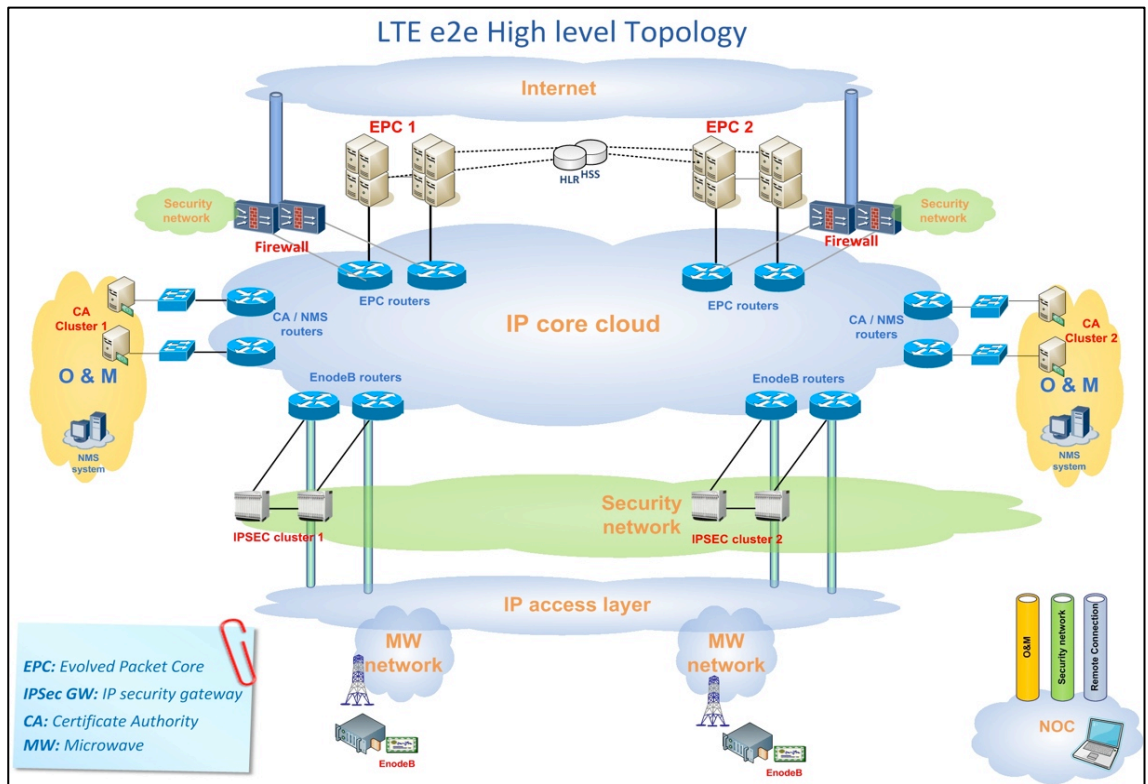

Figure 1 Typical LTE network end-to-end topology

Table 1: LTE Link Budgets

| LTE Link Budget | | | | | | | | | |
|---|---|---|---|---|---|---|---|---|---|
| Morphology | Dense Urban | | Urban | | Suburb | | Rural | | Formulas |
| Data Channel Type | PUSCH | PDSCH | PUSCH | PDSCH | PUSCH | PDSCH | PUSCH | PDSCH | Physical UL/DL Shared Channel |
| Duplex Mode | FDD | | FDD | | FDD | | FDD | | FDD: Frequency Division Duplexing |
| User Environment | Indoor | | Indoor | | Indoor | | Indoor | | |
| System Bandwidth (MHz) | 20.0 | | 20.0 | | 20.0 | | 20.0 | | |
| Channel Model | ETU 3 | | ETU 3 | | ETU 120 | | EVA 120 | | |
| MIMO Scheme | 1×2 | 2×2 SFBC | 1×2 | 2×2 SFBC | 1×2 | 2×2 SFBC | 1×2 | 2×2 SFBC | SFBC: Space-Frequency Block coding |
| Cell Edge Rate (kbps) | 128.00 | 512.00 | 128.00 | 512.00 | 128.00 | 512.00 | 128.00 | 512.00 | |
| MCS | QPSK 0.20 | QPSK 0.12 | QPSK 0.20 | QPSK 0.12 | QPSK 0.20 | QPSK 0.12 | QPSK 0.20 | QPSK 0.12 | QPSK: Quadrature Phase Shift Keying |
| **Tx** | | | | | | | | | |
| Max Total Tx Power (dBm) | 23.00 | 46.00 | 23.00 | 46.00 | 23.00 | 46.00 | 23.00 | 46.00 | A |
| Allocated RB | 3 | 19 | 3 | 19 | 3 | 19 | 3 | 19 | B |
| RB to Distribute Power | 3 | 100 | 3 | 100 | 3 | 100 | 3 | 100 | C |
| Subcarriers to Distribute Power | 36 | 1200 | 36 | 1200 | 36 | 1200 | 36 | 1200 | D = 12*C |
| Subcarrier Power (dBm) * | 7.44 | 15.21 | 7.44 | 15.21 | 7.44 | 15.21 | 7.44 | 15.21 | E = A-10*Log10(D) |
| Beamforming Gain | 0.00 | 0.00 | 0.00 | 0.00 | 0.00 | 0.00 | 0.00 | 0.00 | F |
| Tx Antenna Gain (dBi) | 0.00 | 17.00 | 0.00 | 17.00 | 0.00 | 17.00 | 0.00 | 17.00 | G |
| Tx Cable Loss (dB) | 0.00 | 0.50 | 0.00 | 0.50 | 0.00 | 0.50 | 0.00 | 0.50 | H |
| Tx Body loss (dB) | 0.00 | 0.00 | 0.00 | 0.00 | 0.00 | 0.00 | 0.00 | 0.00 | I |
| EIRP per Subcarrier (dBm) | 7.44 | 31.71 | 7.44 | 31.71 | 7.44 | 31.71 | 7.44 | 31.71 | J = E+F+G-H-I |
| **Rx** | | | | | | | | | |
| SINR (dB) | -4.19 | -5.37 | -4.19 | -5.37 | -2.33 | -4.94 | -2.20 | -4.43 | K |
| Rx Noise Figure (dB) | 2.30 | 7.00 | 2.30 | 7.00 | 2.30 | 7.00 | 2.30 | 7.00 | L |
| Receiver Sensitivity (dBm) | -134.13 | -130.61 | -134.13 | -130.61 | -132.26 | -130.18 | -132.14 | -129.67 | M = K+L-174+10*Log10(15000) |
| Rx Antenna Gain (dBi) | 17.00 | 0.00 | 17.00 | 0.00 | 17.00 | 0.00 | 17.00 | 0.00 | N |
| Rx Cable Loss (dB) | 0.50 | 0.00 | 0.50 | 0.00 | 0.50 | 0.00 | 0.50 | 0.00 | O |
| Rx Body loss (dB) | 0.00 | 0.00 | 0.00 | 0.00 | 0.00 | 0.00 | 0.00 | 0.00 | P |
| Target Load | 75.00% | 90.00% | 75.00% | 90.00% | 75.00% | 90.00% | 75.00% | 90.00% | |
| Interference Margin (dB) | 0.89 | 2.72 | 0.89 | 2.72 | 1.46 | 3.13 | 2.71 | 3.74 | Q |
| Min. Signal Reception Strength (dBm) | -149.74 | -127.89 | -149.74 | -127.89 | -147.31 | -127.05 | -145.93 | -125.93 | R = M-N+O+P+Q |
| **Path Loss & Cell Radius** | | | | | | | | | |
| Indoor Penetration Loss (dB) | 19.00 | 19.00 | 15.00 | 15.00 | 11.00 | 11.00 | 8.00 | 8.00 | S |
| Std. Dev. of Shadow Fading (dB) | 11.70 | 11.70 | 9.40 | 9.40 | 7.20 | 7.20 | 6.20 | 6.20 | |
| Area Coverage Probability | 95.00% | 95.00% | 95.00% | 95.00% | 95.00% | 95.00% | 90.00% | 90.00% | |
| Shadow Fading Margin (dB) | 9.43 | 9.43 | 8.04 | 8.04 | 5.99 | 5.99 | 1.87 | 1.87 | T |
| Maximum Allowable Path Loss (dB) | 128.74 | 131.16 | 134.13 | 136.56 | 137.76 | 141.77 | 143.50 | 147.77 | U = J-R-S-T |
| Propagation Model | Modified Cost231-Hata | | Modified Cost231-Hata | | Modified Cost231-Hata | | Modified Cost231-Hata | | |
| eNodeB/UE Antenna Height (m) | 25.00 | 1.50 | 30.00 | 1.50 | 40.00 | 1.50 | 50.00 | 1.50 | |
| Frequency (MHz) | 1800 | 1800 | 1800 | 1800 | 1800 | 1800 | 1800 | 1800 | |
| Cell Radius (km) | 0.47 | 0.55 | 0.87 | 1.02 | 2.13 | 2.78 | 5.64 | 7.54 | |

* Subcarrier power is estimated assuming that the TX power is equally distrusted across the total bandwidth.

In order to validate the theoretical LB, a field measurement is obligatory. Figure 2 illustrates the UL and DL throughputs versus the pathloss from field results for LTE system at dense urban terrain with similar parameters as in Table 1 [13]. It is evident that the LTE system is UL link limited and the difference in path losses is almost similar to the theoretical one (3-4 dB). After validating the LB, the cell dimensioning process can be conducted as illustrated in the next section.

One of the important factors that may impact the cell radius is the loading. Unlike the legacy universal mobile telecommunications system (UMTS) system, it is not expected that the LTE system will be severely degraded with the increase of the loading thanks to OFDM technique. Since the LB has been validated, we can estimate the cell radius based on different loading scenarios to analyze the impact of the loading on the cell radius. Figure 4, provides the theoretical DL/UL cell radii versus loading for an LTE system at urban terrain with same parameters of Table 1. The figure reconfirms that the LTE is UL limited. The UL cell radius reduction is about 10% at 100% load and it is 5% at 50% load (i.e., typical practical UL load), which is a very graceful degradation compared to HSPA+ system as illustrated later.

For the sake of comparative analysis, the same exercise is conducted with HSPA+ system at 2.1GHz band (i.e., the most popular band for UMTS system) [8], [9]. Figure 4 illustrates the loading impact on the cell radius for UL and DL of HSPA+ system. The figure indicates that the DL is the limiting link in HSPA+ as long as the UL loading is less than 90%. The DL cell radius get decreased by 45% at 100% load while the UL cell radius goes close to 0 at 100% UL load [14]. This illustrates the well know cell breathing drawback of the UMTS system. To further analyze the cell breathing, the cell radius is estimated versus the allocated power of the PA per user to guarantee 512Kbps DL throughput at cell edge. Figure 5 demonstrates the cell radius and number of users with 512Kbps at cell edge versus power allocation per user. The figure indicates that the cell radius shrinks as a function of number of users at cell edge with 512 Kbps throughput and only max of 6 users can achieve 512 Kbps throughput at cell edge while the cell radius shrunk to 200m. Therefore, one native HSDPA carrier (i.e., only HSDPA data) with 40 watt power amplifier (PA) can serve only six users with 512kbps at cell edge, which is only 200 meter away from the cell center [8], [9]. This limitation is a major differentiator for the LTE system thanks to the OFDM technique.

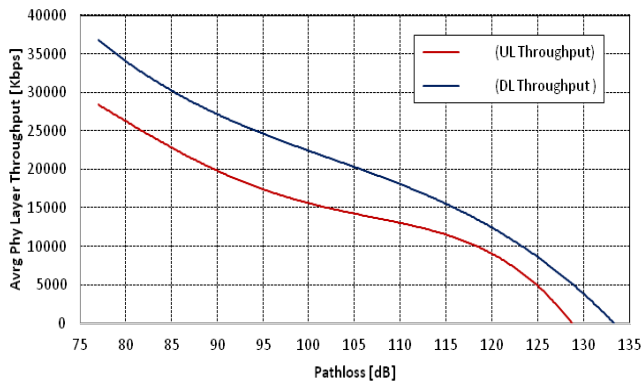

Figure 2 Average layer DL/UL throughputs for LTE system versus pathloss

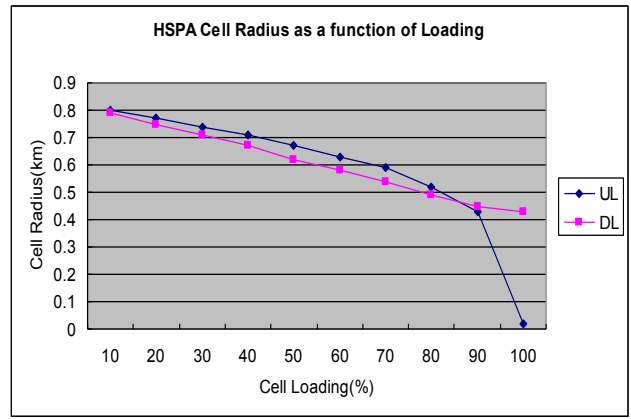

Figure 4 Impact of cell loading on cell radius for HSPA+ at 2.1GHz in urban indoor scenario at 128kbps/512kbps cell edge throughputs for UL/DL, respectively

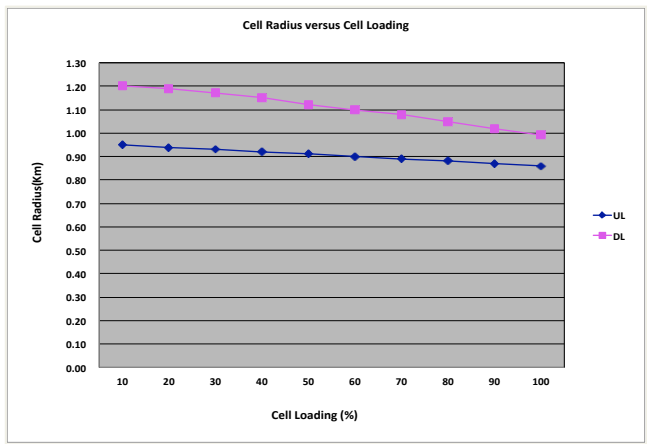

Figure 3 Impact of cell loading on cell radius for LTE system at 1800MHz band with urban indoor scenario at 128kbps/512kbps cell edge throughputs for UL/DL, respectively.

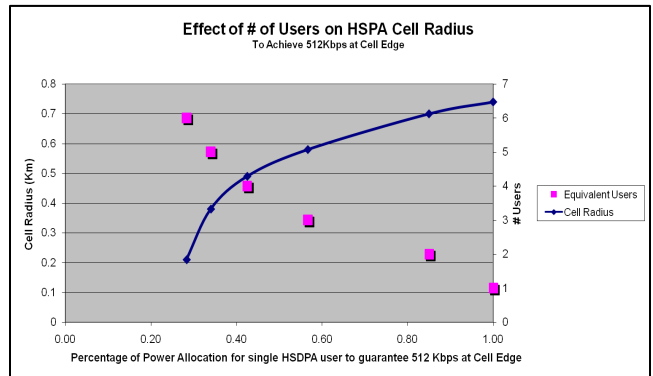

Figure 5 Cell radius and number of users with 512Kbps DL thrghouput at cell edge versus power allocation per user

## IV. LTE Throughput Analysis

For a TCP/IP application and using the generalized protocol parameters in [2], [13], the expected peak theoretical DL and UL throughputs of LTE are provided in Figure 6 for MIMO 2x2 and for single-input single output (SISO) [2]. The SISO throughputs are the expected peak throughputs to be seen in the indoor distributed antenna solution (DAS) that deployed inside tall buildings or underground stations/tunnels to enhance the coverage. Unfortunately, the deployed legacy DAS systems do not support MIMO configuration as only one antenna is deployed and therefore, it is difficult and very costly to upgrade the legacy DAS systems to support MIMO as a complete parallel set of RF feeders and antenna system need to be installed. Considering the throughput gain of MIMO, an alternative solution to deploy MIMO with DAS system based on hotspot approach could be an efficient interim strategy. Another alternative solution is by alternatively feeding the two RF ports of the MIMO 2x2 to feed neighbor DAS antennas, which would provide less degraded MIMO performance. However, DAS design can be revisited to improve MIMO performance by maintaining multiple of half wavelength separation distance between MIMO antennas.

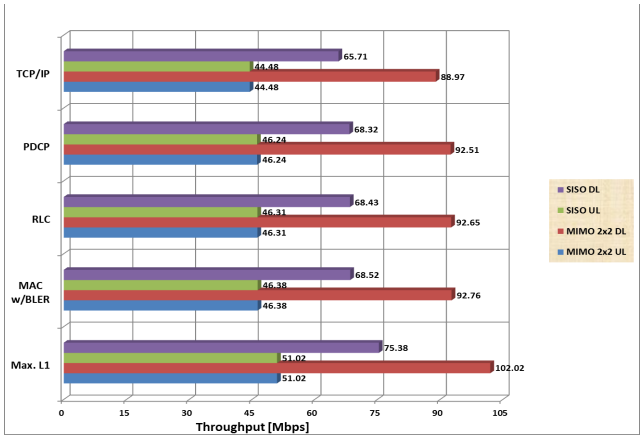

Figure 6 Theoretical LTE UL and DL throughputs for MIMO 2x2 and SISO

The most important factor during network design is the average sector throughput. In this paper we have estimated the average throughput for LTE system based on a commercially deployed LTE network with CAT 3 UE. Figure 7 provides drive test throughput performance of LTE system from a commercially deployed LTE network with same simulation parameters used to estimate the peak throughputs in Figure 6 using MIMO 2x2 configuration. The average throughput is estimated to be around 33Mbps over the entire route with mobility at 80Km/hour. Without losing the generality, we assume the average sector throughput equals the average user throughput (i.e., 33Mbps).

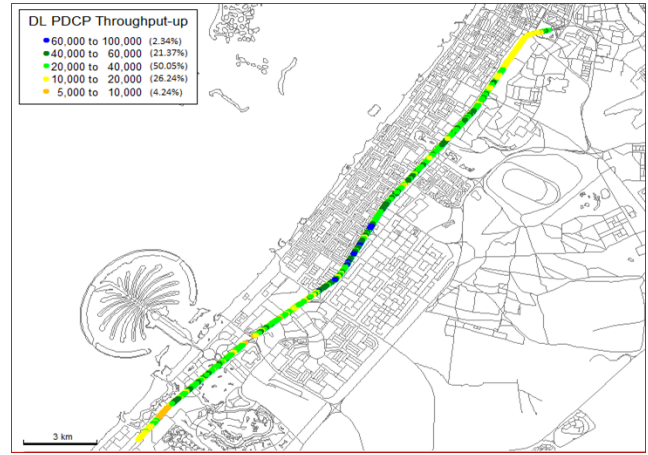

Figure 7 LTE network DL throughput using CAT 3 Modem

## V. LTE Network Dimensioning

Following the average sector throughput estimation as provided in the previous section, the capacity of a single LTE site can be calculated. Figure 8 provides LTE dimensioning exercise and a comparison with DC-HSPA+ system. The key input to the dimensioning exercise is the average sector throughput and the outcome is the number of the subscriber based on the provided traffic profile. The average sector throughput of LTE is 33Mbps as estimated in the previous section. The average sector throughput of the HSPA+ system is estimated using a similar manner to be 12.3Mbps based on collocated LTE/HSPA+ sites and testing same route depicted in Figure 7. A complete comparative comparison between the collocated LTE and HSPA+ systems for the route in Figure 7 is summarized in Table 2 [13]. A typical DL loading of 70% is used in the dimensioning exercise in Figure 8 and the traffic profile of the user is 50kbps during busy hour (i.e., typical user throughput if all users accessed the system at the same time). For sake of comparative analysis, the HSPA+ system is considered with two native HSPA+ carriers (i.e., no R99 traffic) and using the DC-HSPA+ feature where two carriers are aggregated to provide 42Mbps peak throughput. A peak to average margin of 20% is considered to accommodate burst traffic. Figure 8 indicates that the LTE system offers capacity improvement of 34%, which is directly reflected by the average sector throughput gain depicted in Table 2 and thus the spectrum efficiency gain. The complete network dimension exercise is summarized in Figure 9. The full coverage and capacity dimensioning exercise of LTE system is demonstrated in Figure 9 which provides the total number of the required eNBs to meet the coverage and capacity requirements.

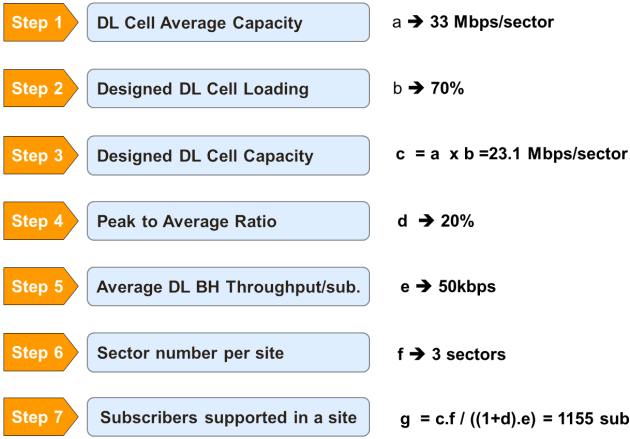
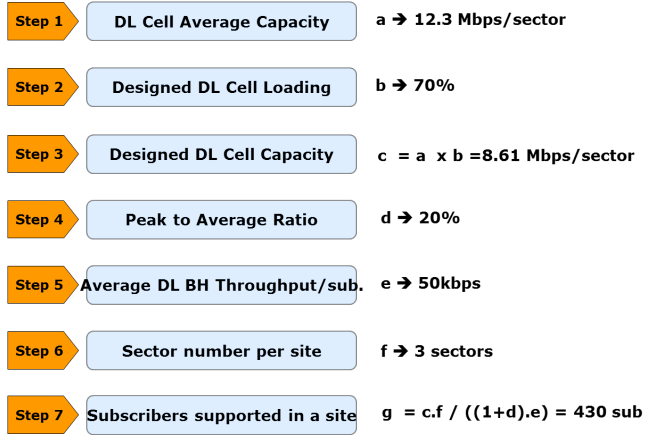

Figure 8 LTE and HSPA+ networks dimensioning

Table 2: Comparison between LTE and HSPA+ based on commercially deployed networks

| Criteria | DC-HSPA+ (2.1GHz) | LTE (1800MHz) |
| --- | --- | --- |
| Mobility average throughput | 9 Mbps with DC (2x5MHz) | 33 Mbps with 20MHz channel BW |
| Average scheduling rate | 73%* | 100% |
| Normalized mobility average Throughput | 12.3 Mbps with DC (2x5MHz) | Same as above |
| Mobility spectrum efficiency | 1.23 (i.e., 12.3Mbps/10MHz) | 1.65 (i.e., 33 Mbps/20Mhz) |
| Throughput % | 2.1% of the route > 21Mbps | 50% of the route > 28Mbps |
| Number of serving cells | 100 | 67 |
| Estimated cell radius (meter) | 390 | 500 (28% improvement) |
| 64QAM utilization % | 8% of the route | 40% of the route |
| MIMO usage % | MIMO +DC is not available yet | 62% of the route |

*Estimated from the number of successful high speed-shared control channel (HS-SCCH) decoded by the UE*

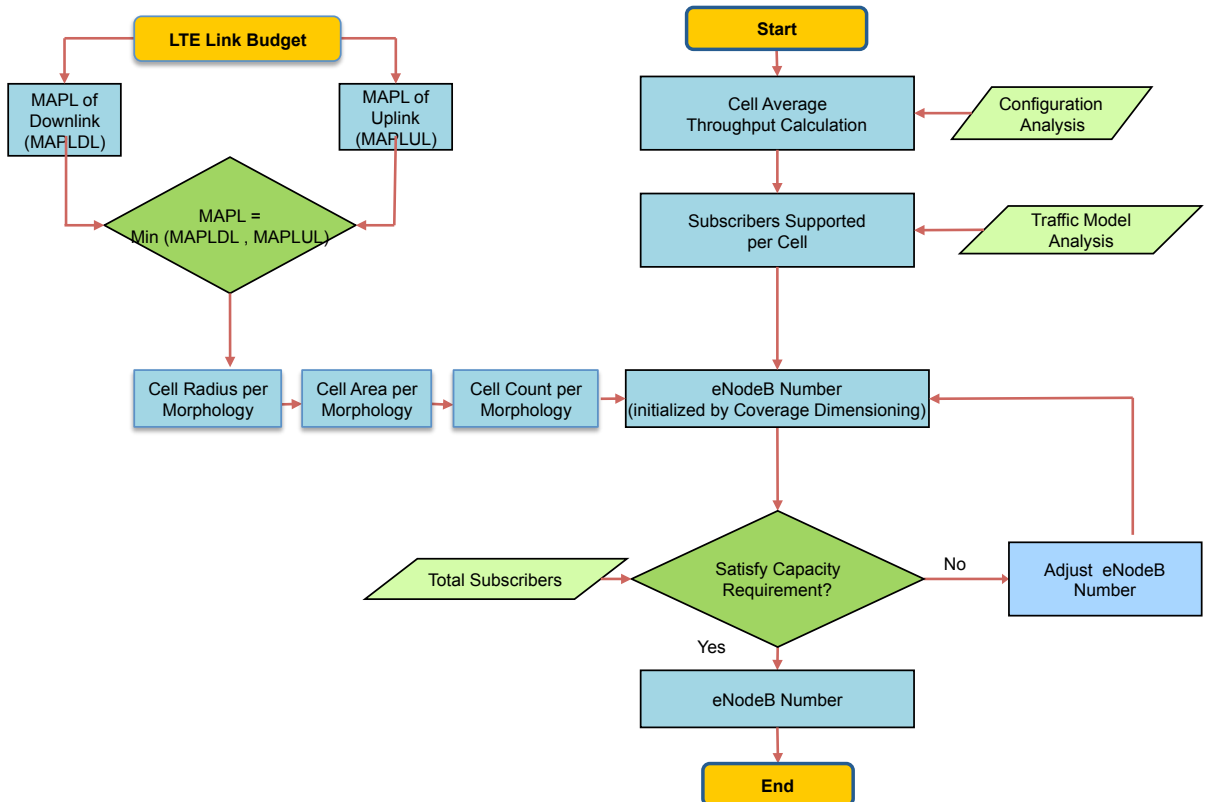

Figure 9 LTE network dimensioning flowchart

## VI. LTE QoS Aspects

The quality of service (QoS) indicates the expected service class in terms of packet delay tolerance, acceptable packet loss rates, and required minimum bit rates during network communication. QoS ensures that request and response of a user (i.e., inter-user QoS) or application (i.e., intra-user QoS) correspond to a certain predictable service class. The QoS is a general term that used on various conditions with service supplies and demands to assess the capability of meeting customer service requirements. The assessment is not based on accurate scoring, but on analysis of service quality in different conditions instead. Then, specific measures can be taken to improve service quality. The most common problems of the IP-based transmission networks are packet delay, jitter, and packet loss. Additionally, different applications require different bandwidths. Therefore, the problem becomes more severe and a robust QoS mechanism is mandatory. The QoS provides a comprehensive solution in such situation. Figure 10 illustrates the end-to-end bearer service architecture.

An evolved packet system (EPS) bearer uniquely identifies packet flows that receive a common QoS treatment between the UE and the PGW (i.e., same scheduling, queue, management/rate, shaping, and policy). As shown in Figure 10, the EPS bearer consists of the radio bearer between the UE and the eNB, the S1-bearer between the eNB and the SGW, and the S5/S8 bearer between the SGW and PGW. An EPS bearer can be a guaranteed bit rate (GBR) or non-guaranteed bit rate (Non-GBR) [2]. Table 3 provides the 3GPP QoS class identifiers (QCI) for different applications with corresponding QoS requirements. The QCI is further used within the LTE access network to define the control packet-forwarding treatment from an end-to-end perspective [11], [12]. It also ensures a minimum standard level of QoS to ease the interworking between the LTE networks mainly in roaming cases and in multi-vendor environments. The packet delay budget (PDB) defines an upper bound delay that a packet is allowed to experience between the UE and the PGW.

The key QoS parameters attached to a bearer are outlined as follows:

1. *QoS Class Identifier (QCI)*: (for inter/intra-user QoS) is used to control packet-forwarding treatment (e.g. scheduling weights, admission thresholds, queue management thresholds, link layer protocol configuration, etc.), and typically pre-configured by the operator.
2. *Allocation and Retention Priority (ARP)* (for Inter-user QoS): the ARP is stored in the subscriber profile in HSS on a per access point name (APN) basis (at least one APN must be defined per subscriber) and it can take value between 1 – 15 based on the user priority (i.e., gold, silver, and bronze). The primary purpose of the ARP is to decide if a bearer establishment/modification request can be accepted or rejected (i.e., admission control) in case of resource limitation.
3. *Guaranteed Bit Rate (GBR) and Maximum Bit Rate (MBR)* – this parameter is defined for GBR bearer only.
4. *Aggregate Maximum Bit Rate (AMBR)* sums all non-GBR bearers per terminal/APN.

The eNB guarantees the downlink GBR associated with a GBR bearer and enforces the downlink AMBR associated with a group of Non-GBR bearers [2]. In order to maintain the same priority of the QCI over end-to-end implementation, the QCI should be mapped to the IP transport network. A key QoS parameter that used for this purpose is the DiffServ Code Point (DSCP) that encapsulated with each IP packet. A typical mapping between QCI and DSCP is also shown in Table 3. The signaling is mapped to the highest priority (i.e., DSCP = 46) while other applications with different QCIs are mapped accordingly. Furthermore, the QCI/DSCP need to be mapped to the transmission network such as the microwave (MW) links air interface. The mapping to the MW air interface is based on queuing technique and the number of queues depends on the MW manufacturer. If we assumed MW with 8 queues, then the mapping is illustrated also in Table 3. It is recommended to reduce the number of queues and number of DSCP values to reduce the complexity during the implementation and later during network expansion or upgrade.

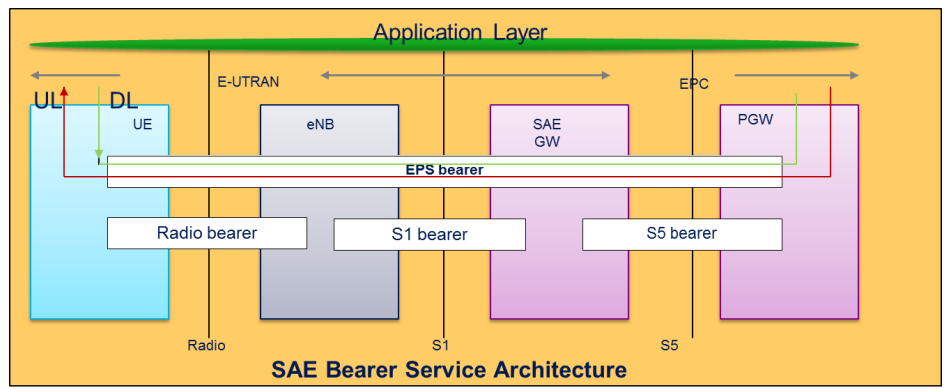

Figure 10 End-to-end SAE bearer service architecture

Table 3. QCI mapping to DSCP and MW queuing

| Traffic type | GBR/Non-GBR | QCI | Priority | Packet Delay Budget | Packet loss rate | DSCP | MW Queuing | Service Sample |
|---|---|---|---|---|---|---|---|---|
| Signaling (SCTP) | N/A | | | | | 46 | 7 | Stream Transmission Control Protocol |
| Sync (1588V2) | | | | | | 46 | 7 | Synchronization Signal |
| O&M | | | | | | 46 | 7 | Operation and maintenance |
| User Data | GBR | 1 | 2 | 100 | $10^{-2}$ | 46 | 7 | Conversational voice |
| | | 2 | 4 | 150 | $10^{-3}$ | 26 | 4 | Conversational video (Live Streaming) |
| | | 3 | 3 | 50 | $10^{-3}$ | 34 | 5 | Real time gaming |
| | | 4 | 5 | 300 | $10^{-6}$ | 26 | 4 | Non-conversational video (Buffer streaming) |
| | Non-GBR | 5 | 1 | 100 | $10^{-6}$ | 46 | 7 | IMS signaling |
| | | 6 | 6 | 300 | $10^{-6}$ | 18 | 2 | Video (buffer steaming) TCP based (www. e-mail, chat, ftp) |
| | | 7 | 7 | 100 | $10^{-3}$ | 18 | 2 | Voice, Video (live streaming) interactive streaming. |
| | | 8, 9 | 8, 9 | 300 | $10^{-6}$ | 0 | 0 | Video (buffer steaming) TCP based (www. e-mail, chat, ftp) |

The deployment scenario in Table 3 is for application-based QoS (i.e., provide different treatment for different applications). This scenario consumes almost all MW capability in terms of queues and therefore, there is no room to accommodate other technologies on the same MW network. Another interesting scenario to deploy QoS is the inter-user QoS by providing differentiated treatment for users based on their importance, for example, gold, silver, and bronze users. This approach will allow the operator to offer differentiated services to their customers while maximizing the utilization efficiency of the scarce resources in air interface and transport network. Table 4 provides a practical deployment scenario for the inter-user QoS in commercial LTE network. In this exercise, QCI 6, 8, and 9 are used to represent gold, silver, and bronze users, respectively. The user priority (i.e., ARP) is stored in the HSS profile of the user and the PGW can map the QCI values to DSCP values as per table 4. The PGW cannot map the ARP to DSCP. The scheduling weight for each user category is defined in the eNB and it is associated with each ARP/QCI. The throughput testing results of this exercise is shown in Figure 11. In this exercise, one LTE cell with 20MHz channel bandwidth is used and three users (gold, silver, and bronze) are accessing the LTE cell at the same time. The demonstrated throughput is obtained using FTP download.

Table 4: Inter-user QoS deployment scenario in commercial LTE network

| User Category | DSCP Values in Transport Network | | | User Priority (ARP) in HSS | QCI | Allocation/Priority Weight in eNB |
|---|---|---|---|---|---|---|
| | Signaling | Voice | Data Traffic | | | |
| GOLD | 46 | 46 | 34 | 5 | 6 | 100 |
| Silver | 46 | 46 | 18 | 7 | 8 | 50 |
| Bronze | 46 | 46 | 0 | 9 | 9 | 20 |

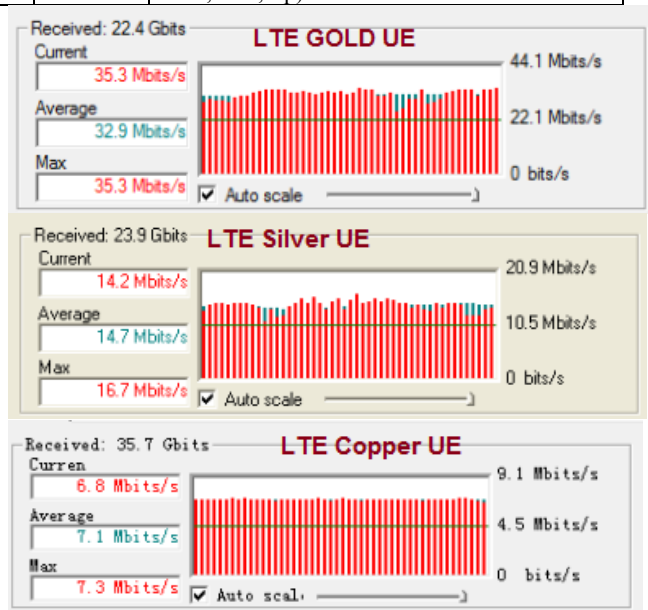

Figure 11 FTP download throughput for three users with different QoS as per Table 4.

As shown in Table 5, the actual average users throughputs are well presented by the allocation/priority weight. The allocation/priority weight is operator adjustable value and it depends on operator strategy and pricing scheme for each user category or the data bundle.

Table 5: Throughput Analysis for the scenario in Table 4

| Total Cell Throughput (Mbps) | 54.7 | | |
|---|---|---|---|
| User Priority | Gold | Silver | Copper |
| Allocation/ Priority Weight in eNB | 100 | 50 | 20 |
| Number of Online Users | 1 | 1 | 1 |
| Expected Average User Throughput (Mbps) | 32.2 | 16.1 | 6.4 |
| Expected Throughput Percentage (%) | 59% | 29% | 12% |
| Achieved Average Throughput (Mbps) | 32.9 | 14.7 | 7.1 |
| Achieved Throughput Percentage (%) | 60% | 27% | 13% |

## VII. LTE network Latency

The latency is one of the most critical factors that can impact the performance of the LTE network. Therefore, a special attention should be given to the latency at designing phase of the LTE network to meet the X1 and S2 latency requirements and also to maintain the gain of the reduced latency that offered by the LTE system. A latency comparison is conducted between the LTE simplified topology (termed as LTE trial) where the eNB is directly connected to the EPC without IPSec GW via one transmission media (i.e., fiber network in this case) and a full commercial LTE network similar to the topology in Figure 2. Also, for sake of comparative analysis, the latency of the collocated HSPA+ network is presented. The latency is measured in terms of round trip time (RTT) which is estimated by sending packets of different sizes (i.e., 10, 100, 1000, 1460 Bytes) using ping command to a local FTP server directly connected to the EPC in case of LTE network and to the packet core of the HSPA+ network (i.e., GGSN; gateway GPRS support node). As demonstrated in Figure 12, the simplified LTE network offers significant latency reduction. Moreover, the LTE system can maintain the same latency even with bigger packet size. The full commercial LTE network introduces higher latency compared to the simplified LTE network due to the introduction of the IPSec GW, firewall, and other IP routers in access and core clouds. Despite this, the full commercial LTE network yields a reduced latency compared to the collocated HSPA+ network that uses the same backhauling. More specifically, the full commercial LTE network offers average reduction of 40% in latency compared to HSPA+ network. Furthermore, with HSPA+ network, the latency is significantly increased with the packet size increase where the difference in latency between the smallest and biggest packets is 21ms versus 12ms with the full commercial LTE network and only 1ms for the simplified LTE network.

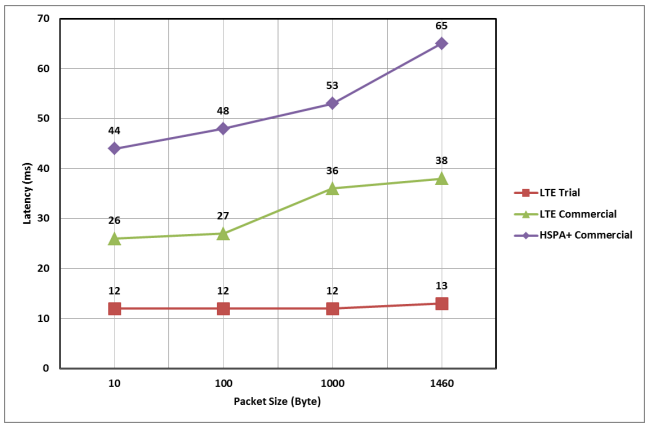

Figure 12 Latency comparison among LTE simplified network, full LTE network, and collocated HSPA+ network

## VIII. Conclusions

In this paper, we have provided the key practical aspects and the best practices for design and deployment of a commercial LTE network. The end-to-end LTE network topology is presented along with key network domains. The theoretical LB of the LTE system is analyzed and validated using field test results. It has been validated that the LTE system is UL link limited and the difference between UL and DL path losses is around 3-4 dB. On the other side, the DL is the limiting link in HSPA+ system as long as the UL loading is less than 90%. Moreover, with LTE system, the UL cell radius is reduced by 10% at 100% load, which is a very graceful degradation compared to the HSPA+ system where the DL cell radius is decreased by 45% at 100% load. We have conducted an end-to-end dimensioning exercise for LTE system including coverage and capacity planning. In addition, we have presented the key practical aspects of the QoS deployment and the end-to-end deployment scenario for a commercial LTE network for both application-based QoS and inter-user QoS. Also, a practical throughput testing results from live LTE network is presented for inter-user QoS. Finally, latency of the LTE network is compared with the HSPA+ network. It has been demonstrated that the full commercial LTE network offers average reduction of 40% in latency compared to HSPA+ network.

## Acknowledgments

I wish to express my appreciation to the wireless broadband team for their cooperative support. Many thanks to Huawei and Qualcomm teams who supported this work with a lot of resources and practical results.

**Ayman Elnashar** was born in Egypt in 1972. He received the B.S. degree in electrical engineering from Alexandria University, Alexandria, Egypt, in 1995 and the M.Sc. and Ph.D. degrees in electrical communications engineering from Mansoura University, Mansoura, Egypt, in 1999 and 2005, respectively. He has more than 15 years of practical experience in telecoms industry including GSM, GPRS/EDGE, UMTS/HSPA+/LTE, WiMax, WiFi, and transport/backhauling technologies. Currently, he is Sr. Director of Wireless Broadband with the Emirates Integrated Telecommunications Co. "du", UAE. He is in charge of mobile and fixed wireless broadband networks.

Prior to this, he was with Mobily, Saudi Arabia, from June 2005 to Jan 2008 and with mobinil (orange), Egypt, from March 2000 to June 2005. He has managed several large-scale networks and mega projects with approximate capital of one billion USD including start-up, networks expansion, and swap projects. He obtained his PhD degree in multiuser interference cancellation and smart antennas for cellular systems. He published 18 papers in wireless communications arena in highly ranked journals such as *IEEE Transactions on Antenna and Propagation*, *IEEE Transactions Vehicular technology*, and *IEEE Transactions Circuits and Systems* and international conferences. His research interests include digital signal processing for wireless communications, performance analysis of cellular systems, CDMA, OFDM, mobile network planning and design, multiuser detection, smart antennas, beamforming, and robust adaptive detection.